# The Central Role of Tilted Anisotropy for Field-Free Spin-Orbit Torque Switching of Perpendicular Magnetization

*Chen-Yu Hu[†], Wei-De Chen[†], Yan-Ting Liu, Chao-Chun Huang, and Chi-Feng Pai\**

Chen-Yu Hu, Wei-De Chen, Yan-Ting Liu, Chao-Chun Huang, and Chi-Feng Pai

Department of Materials Science and Engineering, National Taiwan University

Taipei 10617, Taiwan

E-mail: cfpai@ntu.edu.tw

Chi-Feng Pai

Center of Atomic Initiative for New Materials, National Taiwan University

Center for Quantum Science and Engineering, National Taiwan University

Taipei 10617, Taiwan

Key words: The spin Hall effect, SOT-MRAM, field-free switching, tilted anisotropy, spin-orbit torque, spin-transfer torque, magnetic anisotropy

**Abstract**

The discovery of efficient magnetization switching activated by the spin Hall effect (SHE)-induced spin-orbit torque (SOT) changed the course of magnetic random-access memory (MRAM) research and development. However, for systems with perpendicular magnetic anisotropy (PMA), the use of SOT is still hampered by the necessity of a longitudinal magnetic field to break the magnetic symmetry to achieve deterministic switching. In this work, we first demonstrate that a robust and tunable field-free current-driven SOT switching of perpendicular magnetization can be controlled by the growth protocol in Pt-based magnetic heterostructures. It is further elucidated that such growth-dependent symmetry breaking is originated from the laterally tilted magnetic anisotropy of the ferromagnetic layer with PMA, which has been largely neglected in previous studies and its critical role should be re-focused. We show by both experiments and simulations that in a PMA system with tilted anisotropy, the deterministic field-free switching possesses a conventional SHE-induced damping-like torque feature and the resulting current-induced effective field has a non-linear dependence on the applied current density, which could be potentially misattributed to an unconventional SOT origin.

**Introduction**

The ultrafast and energy-efficient characteristics of employing spin-orbit torque (SOT)[1-3] for magnetization switching make it a promising writing mechanism for next generation magnetic random-access memory (MRAM)[4, 5]. In various types of magnetic heterostructures, by applying an in-plane current ($J_c \parallel \hat{x}$) through the spin current sources (SCSs), such as the bulk materials with sizable spin Hall effect (SHE)[6-8] or the interfaces with strong Rashba effect[9, 10], the generated spin current with transverse polarization ($\sigma \parallel \hat{y}$) would transfer spin angular momentum to the neighboring magnetization ($M$) then activate magnetic dynamics. Such spin-torque-induced dynamics of $M$ can be described by the Landau-Lifshitz-Gilbert-Slonczewski (LLGS) equation[11, 12] comprised of a field-like (FL-) and a damping-like (DL-) SOT term, where $\tau_{FL} \propto M \times \sigma$ and $\tau_{DL} \propto M \times (\sigma \times M)$. Considering the scalability and stability aspects[13], magnetic heterostructures with perpendicular magnetic anisotropy (PMA) are deemed as the best candidates for SOT-MRAM applications. However, limited by the symmetry of SOT[2], either $\tau_{DL}$ or $\tau_{FL}$ cannot provide the necessary symmetry breaking between the UP and DOWN state of the magnetic materials (FM) with PMA, which leads to a non-deterministic SOT switching behavior. To break the symmetry and realize the deterministic SOT switching, the simplest way is to apply an in-plane field along the current direction ($H \parallel \hat{x}$)[14-16], which has been demonstrated by introducing a magnetic hard mask (MHM)[17] in the SOT-MRAM cell with PMA. Nevertheless, a genuine field-free solution is much more desirable for industrial applications.

By far some commonly-studied symmetry breaking mechanisms for achieving field-free SOT switching of a PMA FM layer include: Tilted magnetic anisotropy (anisotropy gradient[18], device shape engineering[19], magneto-crystalline anisotropy[20], mixed magnetic anisotropy[21, 22]), unconventional spin currents (low symmetry materials[23], magnetic spin Hall effect[24]), perpendicular FL-SOT (oxidation gradient at CoFeB/oxide interface[25]), symmetric[26] and antisymmetric[27] interlayer exchange coupling, exchange bias coupling[28, 29], and also the orange peel effect[30]. Besides above-mentioned approaches, oblique angle deposition of magnetic heterostructures is another method to introduce structural symmetry breaking in PMA systems, but the corresponding mechanism for field-free switching strongly depends on the stack design and device fabrication process. For example, the lateral magnetic anisotropy asymmetry from ferromagnetic layer thickness gradient is concluded to be the field-free switching origin for the Pt/Co(obliquely-deposited)/MgO structure[16]; the SOT gradient is for the Cu-Pt(obliquely-deposited)/[Co/Ni]$_n$ structure[31]; and the tilted magnetic anisotropy is for the Pt/Co(obliquely-deposited)/Pt structure[18] and the heavy-metal(obliquely-deposited)/CoFeB(obliquely-deposited)/MgO structure[32]. Due to these diverse proposed mechanisms and the lack of a universal and engineering-friendly protocol to realize robust field-free SOT switching, we therefore try to scrutinize the origin of symmetry breaking in a simple PMA system, namely Pt/Co/Pt trilayer, as prepared by different oblique deposition approaches. Our experimental and simulated results suggest that tilted magnetic anisotropy is strongly correlated to the oblique deposition angle and the texture transformation of the Pt buffer layer. Field-free switching in such system can also be explained by a conventional SHE-induced DL-SOT plus tilted magnetic anisotropy scenario, in which the inclusions of unconventional spin current or exotic SOT terms are unnecessary.

**Materials and methods**

**Samples preparation**

We sputter-deposited Ta(0.5)/Pt(5)/Co(0.5)/Pt(1)/Ta(2) (units in nanometers) with PMA on thermally-oxidized Si substrates. The bottom Ta(0.5) serves as a seed layer to ensure the uniformity of the films, and the top Ta(2) is expected to be naturally oxidized to prevent other layers from oxidation. The devices were prepared in a magnetron sputtering system with base pressure of $1 \times 10^{-8}$ Torr and working pressure of 3 mTorr by Ar gas flow. Hall bar devices with lateral dimensions of 5 μm × 60 μm and 10 μm × 60 μm were patterned subsequently for electrical measurements. The oblique angle deposition in this work is achieved by using a wedge-shaped sample holder with different oblique angles, and is controlled by the deposition angle ($\theta_{dep}$) of the Pt atom flow based on our chamber design. The azimuthal angle of deposition ($\alpha$) is controlled by the placement of Hall bar pattern on the wedge-shaped holder, as shown in **Fig.** 1(a). For instance, based on the coordinate of the Hall bar device $(x, y, z)$, the Pt atom flow incomes from the $-y$ direction of the Hall bar device for the $\alpha = 0°$ geometry, and from the $-x$ direction for the $\alpha = 90°$ geometry, and so on. For the control samples prepared with $\theta_{dep} = 0°$, the sample holder rotates during the whole deposition process to avoid wedge effects. As for those under $\theta_{dep} = 20°, 30°, 40°$, the sample holder remains static during the deposition of Ta(0.5)/Pt(5) buffer layer, for controlling the direction of the incoming atom flow. The sample holder rotation is resumed for depositing the rest of the layers (Co(0.5)/Pt(1)/Ta(2)). The sputtering rate of each material is respectively calibrated for different $\theta_{dep}$ conditions by a profilometer.

**Basic electrical and magnetic properties**

Electrical measurements on Hall bar devices are performed with a current source (Keithley 2400) and a voltage meter (Keithley 2000) integrated into a probe station with or without electromagnets based on the characterization purpose. The electromagnets are typical home-made coils with iron cores in the case applying typical field along $x, y, z$-axis, and a commercial projected vector field magnet of model GMW Associates Model 5204 for applying arbitrary vector fields based on the previously established framework[33]. An X-ray diffractometer (XRD) of model Rigaku TTRAX 3 is used to characterize the crystallinity of the deposited films. A vibrating sample magnetometer (MicroSense EZ9 VSM) is used to characterize the saturation magnetization and the dead layer thickness.

Longitudinal resistivities ($\rho_{xx}$) of the Pt and the Co layers are obtained by four-points probe measurement and the parallel resistor model, giving $\rho_{Co} = 85.32\ \mu\Omega \cdot cm$ and $\rho_{Pt} = 18.8\ \mu\Omega \cdot cm$, and the deposition angle shows a negligible influence on the film resistivity. The VSM characterization results show that the Co layer has a saturation magnetization of $M_s = 1132.3 \pm 30.9\ emu/cm^3$ with a negligible dead layer thickness ($t_{dead} \sim 0$ nm). The PMA properties of the magnetic heterostructures prepared under different $\theta_{dep}$ are evaluated via the typical anomalous Hall effect (AHE) hysteresis measurement[34], as shown in **Fig.** 1(b). The coercivity ($H_c$) is defined as when the normalized Hall resistance ($R_H$) changes its sign as sweeping an out-of-plane magnetic field ($H_z$); and the ratio between the remnant perpendicular magnetization under zero external field and the saturated magnetization is defined by the normalized $R_H$ under $H_z = 0$ Oe ($M_r/M_s \equiv R_H^{H_z=0\ Oe}/R_H^{max}$), and is used as a figure-of-merit for evaluating PMA. The out-of-plane $H_c$ and $M_r/M_s$ for each $\theta_{dep}$ condition are summarized in **Fig.** 1(c), showing that the $H_c$ (ranging from 30 to 40 Oe) and the PMA degree are non-sensitive to the varying $\theta_{dep}$ in these Hall bar devices.

**Results**

**Demonstration of robust field-free switching in obliquely deposited Pt/Co/Pt structures**

The experimental setup for field-free switching is schematically shown in **Fig.** 2(a). The switching sequence consists of alternatively applying write current pulse ($I_{wrte}$) of various amplitudes and read current pulse ($I_{read}$) of 0.1 mA into the Hall bar devices to alter and read out the magnetization state through AHE. The pulse width is set as 0.05 s. The experiments were conducted in an external-field-free condition (probe station without electromagnets). **Fig.** 2(b) shows representative field-free switching loops for the devices prepared under $\theta_{dep} = 0°, 40°$ and $\alpha = 0°, 180°$. For the case of $\theta_{dep} = 0°$, both $\alpha = 0°$ and $180°$ devices are not capable to achieve full switching ($R_H$ difference in cases of field-sweeping and current-sweeping are nearly identical) under field-free condition. However, full switching loops with opposite polarities can be observed in the case of $\alpha = 0°, 180°$ prepared under $\theta_{dep} = 40°$. To confirm the deterministic characteristics of these field-free switching loops, we further characterized the switching probabilities ($P_{sw} \equiv \frac{\#succesful\ switching\ events}{\#total\ switching\ events}$) of the devices prepared under $\alpha = 0°$ and different $\theta_{dep}$. Within a single tentative switching event, a sequence of $+I_{write}, +I_{read}, -I_{write}, +I_{read}$ pulses are applied into the device. A successful switching event is defined as when the corresponding normalized $R_H$ is over $\pm 0.7$ after applying $\pm I_{write}$, as represented by the colored regions in **Fig.** 2(c). $P_{sw}$ (per 100 switching events) vs. $I_{write}$ for devices with various $\theta_{dep}$ are summarized in the **Fig.** 2(d). The zero-switching probability for the $\theta_{dep} = 0°$ device indicates that non-wedge structure cannot realize field-free switching, as expected. In contrast, devices prepared by oblique angle deposition can achieve 100% $P_{sw}$. Furthermore, a clear reduction in $I_{write}$ to reach full deterministic switching is observed by increasing $\theta_{dep}$: 14.4 mA for $\theta_{dep} = 20°$, 11.6 mA for $\theta_{dep} = 30°$, and 6.8 mA for $\theta_{dep} = 40°$. This implies that the spin torque response enforcing on the Co layer can be tuned by adjusting $\theta_{dep}$, which could be originated from the change of magnetic properties (such as a tilted easy axis of the magnetization) or the change in spin current generation mechanism (structure-dependent unconventional spin current). Note that by using a two-step fabrication process, we found that the oblique angle deposition of the bottom Pt layer, while compared to other layers, plays a more decisive role of realizing field-free current-induced switching (see Supporting Information S1).

**Evaluation on the tilted magnetic anisotropy**

We first examined the possibility of having tilted easy axes in these samples with non-zero $\theta_{dep}$. To characterize such potentially minute change in the easy axis direction, we adopted the modified AHE hysteresis loop shift measurements with an in-plane magnetic field $H_{IP}$ scanning along various azimuthal angles $\varphi_H$ [18, 22], as shown in **Fig.** 3(a). The working principle of this measurement is that if the easy axis tilts away from the z-axis, the non-zero component of the applied external in-plane field projected onto the easy axis would lead to a shift in the out-of-plane hysteresis loop while sweeping $H_z$; on the contrary, no shift will be observed in the case of easy axis lying exactly on the z-axis. To prevent sizable contributions from the Joule heating and spin torque effects, the longitudinal sense current to generate Hall voltage is set as 0.1 mA.

Representative AHE data obtained from a $\theta_{dep} = 40°$ device is shown in **Fig.** 3(b), from which an obvious shift in the out-of-plane hysteresis loop (denoted as $\Delta H_{sw}$) is observed under $H_{IP} = 750$ Oe and $\varphi_H = 90°$.

The exact direction of the tilted easy axis, which is parameterized by $\theta_{ani}$ and $\varphi_{ani}$ (the polar and the azimuthal angle), can be estimated from the loop shift results as follows. First, the sum of the externally applied in-plane fields along the easy axis ($H_{ea}$) is:

$$H_{ea} = H_z \cos\theta_{ani} + H_{IP} \cos(\varphi_{ani} - \varphi_H) \sin\theta_{ani}. \tag{1}$$

Since the switching fields along the easy axis should be the same for DOWN-to-UP and UP-to-DOWN transitions ($H_{ea}^{DN-to-UP} = -H_{ea}^{UP-to-DN}$), the shift of the AHE hysteresis loop center $\Delta H_{sw} \equiv \frac{1}{2}(H_{sw}^{UP-to-DN} + H_{sw}^{DN-to-UP})$ obtained experimentally is therefore linearly proportional to $H_{IP}$:

$$\frac{\Delta H_{sw}}{H_{IP}} \equiv \frac{H_z^{DN-to-UP} + H_z^{UP-to-DN}}{2H_{IP}} = -\cos(\varphi_{ani} - \varphi_H)\tan\theta_{ani}, \tag{2}$$

where $H_z^{DN-to-UP}$ and $H_z^{UP-to-DN}$ respectively are the switching fields extracted from the AHE hysteresis loops. Such linear relationship under different $\varphi_H$ for a $\theta_{dep} = 40°$ device is shown in **Fig.** 3(c). The values of $\theta_{ani}$ and $\varphi_{ani}$ then can be extracted by fitting $\frac{\Delta H_{sw}}{H_{IP}}$ vs. $\varphi_H$ using Eqn. (2), as shown in **Fig.** 3(d).

We summarize the oblique deposition angle $\theta_{dep}$ dependence of the tilted easy axis ($\theta_{ani}$, $\varphi_{ani}$) in **Fig. 3(e)**. The negligible $\theta_{ani}$ and $\varphi_{ani}$ for the control sample prepared under $\theta_{dep} = 0°$ indicates a typical PMA behavior that the easy axis is fairly close to the z-axis ($\theta_{ani} = 0.74°$ and $\varphi_{ani} = 14.57°$). For the devices prepared by oblique angle deposition, $\theta_{ani}$ gradually increase as increasing $\theta_{dep}$ ($\theta_{ani} = 5.32°$ for $\theta_{dep} = 40°$). The easy axis also gradually points toward y-axis as using larger $\theta_{dep}$ ($\varphi_{ani} = -80.73$ for $\theta_{dep} = 40°$). The origin of this tilted anisotropy is tentatively attributed to the tilted (111) texture observed in the Pt films prepared by oblique angle deposition, which is quite different from the normal (111) texture observed from the $\theta_{dep} = 0°$ films (see Supporting Information S2). Note that this mechanism is different from the one attributed to the Co thickness gradient[18], since the Co layer in this work was grown uniformly. Also note that the polycrystalline nature of the sputter-deposited Pt also rules out the possibility of producing any sizable unconventional spin currents related to the broken crystal symmetry[35].

**Current-driven field-free SOT switching: Conventional DL-SOT plus tilted anisotropy**

To further investigate the origin of field-free current-induced switching under the existence of tilted anisotropy, current-induced AHE hysteresis loop shift measurements were conducted in an in-plane-field-free probe station, as illustrated in **Fig. 4(a)**. Via this approach, we can obtain the out-of-plane AHE loop shift $\Delta H_{sw} \equiv \frac{1}{2}(H_{sw}^{UP-to-DN} + H_{sw}^{DN-to-UP})$ as a function of applied current ($I_{applied}$), ranging from $-9.5$ mA to $+9.5$ mA in the absence of $H_{IP}$. Representative current-induced AHE loop shift results for a $\theta_{dep} = 40°$ device under $I_{applied} = \pm 7.5$ mA is shown in **Fig. 4(b)**. A clear DL (anti-damping) torque switching behavior is identified: the SOT only affects one of the magnetization switching fields $H_{sw}$, either that for DOWN-to-UP or UP-to-DOWN transition. Such behavior is not observed in the control sample with $\theta_{dep} = 0°$ (Details of the switching fields of $\theta_{dep} = 0°$ and $\theta_{dep} = 40°$ under different $I_{applied}$ can be found in Supporting Information S3).

More importantly, the AHE loop shifts $\Delta H_{sw}$ vs. $I_{applied}$ as summarized in **Fig.** 4(c) show clear non-linear trends: $\Delta H_{sw}$ is negligible in the small current regime, and a sizable $\Delta H_{sw}$ can only be observed when the applied current is beyond a threshold value. This feature indicates that a large current is required to activate the anti-damping torque for field-free switching, and can be seemed as an evidence of the existence of spin polarization component partially parallel to the magnetization, similar to the case regarding a spin current with $z$-spin polairzation[24]. However, in our case, the parallel amount is related to the anisotropy tilt toward $y$-axis, where $\sigma \cdot \hat{m} = \sigma \sin \theta_{ani}$ per spin angular momentum ($\sigma$). The spin polarization here is still conventional, *i.e.*, $y$-spin. It is noteworthy that this non-linear torque response has already been observed and speculated as a result of tilted anisotropy in a previous work[32], but the existence of a tilted anisotropy and the switching field feature of a conventional DL torque behavior have not been fully addressed and verified. Here we unambiguously demonstrate that a conventional $y$-spin-polarization-induced DL-SOT plus tilted anisotropy scenario should be sufficient to explain the observed field-free switching.

The proposed mechanism is further verified via microspin simulation based on UberMag package[36] (with OOMMF[37] as kernel) to model current-induced hysteresis loop shifts $\Delta H_{sw}$ under different applied current densities and anisotropy conditions. The modeled device size is set as $720 \times 120 \times 1$ nm$^3$ and is discretized by using a cuboid-shaped cell of $4 \times 4 \times 1$ nm$^3$ with uniform anisotropy energy of $8 \times 10^5$ J/cm$^3$, DMI energy of $1$ J/m$^2$ and exchange energy of $16 \times 10^{-12}$ J/m. The spin Hall angle ($\theta_{SH}$) is set to be 0.2 with a conventional spin polarization along $y$-axis. As shown in **Fig. 4(d)**, $\Delta H_{sw}$ remains zero under various current densities $J_{applied}$ for the $\theta_{ani} = 0°$ configuration, which is consistent with the fact that the net SOT should be zero without the application of a longitudinal in-plane magnetic field. In the case with a laterally tilted anisotropy ($\theta_{ani} = 5°$, $\varphi_{ani} = -90°$), a clear threshold behavior of $\Delta H_{sw}$ vs $J_{applied}$ agrees well with the above-mentioned experimental results and again verifies the non-linear DL torque response in a system with tilted anisotropy. Unconventional spins along $z$-direction is not necessary to be included for reproducing such non-linear feature.

Additionally, as presented in **Fig. 4(e)**, such non-linear $\Delta H_{sw}$ vs $I_{applied}$ behavior can be controlled by the direction of oblique angle deposition α. The α-dependence of $\Delta H_{sw}$ shown in **Fig. 4(f)** also suggests that the largest $\Delta H_{sw}$ happens when using oblique angle deposition along the transverse direction (α = 0°), while $\Delta H_{sw} \approx 0$ Oe for the longitudinally-deposited device (α = 90°). A clear sinusoidal trend is observed (absent) as in the case of large (small) applied current, implying that the Pt(111) texture (and the corresponding tilted anisotropy direction) controlled by $\theta_{dep}$ and the sufficient current for anti-damping torque activation are both important for the observation of this non-linear response in the absence of in-plane field.

**Longitudinal effective in-plane field in the presence of tilted anisotropy**

Finally, we examine and compare the difference of the field-dependent SOT responses between devices with and without tilted anisotropy, by both conventional current-induced hysteresis loop shift and current-induced SOT switching measurements in the presence of an externally applied in-plane field. The applied current in the hysteresis loop shift measurement is set at ±6.5 mA to reflect the behavior within activation regime, and the corresponding results are summarized in **Fig.** 5(a, b). For the case without tilted anisotropy, a classical trend[16] of perpendicular effective field per current ($\Delta H_{sw}/I$) is observed as expected, that when applying a longitudinal external field along *x*-axis, $H_x$, the magnitude of $\Delta H_{sw}/I$ starts to increase from zero then saturates. No observable $\Delta H_{sw}/I$ is shown when applying transverse external field ($H_y$) due to the missing of domain expansion/shrinkage in the Bloch-type-like domain wall geometry. On the contrary, these behaviors become quite different when a tilted anisotropy is introduced into the system: A finite $\Delta H_{sw}/I$ in the zero $H_{x(y)}$ condition is observed for the $\theta_{dep} = 40°$ device (**Fig. 5b**), which contributes to the field-free switching when applying a sufficient amount of current into the device. Moreover, the overall $\Delta H_{sw}/I$ vs. $H_x$ shows a clear shift in the longitudinal direction, and can be suppressed as $H_x$ reaches around 150 Oe, implying the outcome of the laterally tilted anisotropy can be phenomenologically viewed as a negative longitudinal effective field.

This effective field orthogonal to the spin polarization direction and the tilted easy axis can be attributed to the DL-SOT emerges as the magnetization *m* is activated and lay in the hard plane. Without tilted anisotropy, such DL-SOT effective field is proportional to $m \times \sigma = 0$ or parallel to *z*-axis, thereby missing the in-plane field component. However, with a laterally tilted anisotropy, the DL-SOT effective field experiences by the magnetization within the hard plane is non-zero and along $m \times \sigma \parallel \hat{x}$. In our case of having a conventional *y*-axis spin polarization ($\sigma \parallel \mp\hat{y}$) carried by a positive (negative) charge current in the Pt layer (along *x*-axis), *m* lays within the negative (positive) *yz*-plane would both experience a negative DL-SOT effective field pointing along *x*. This characteristic is also reported in a similar system with canted easy axis, albeit having in-plane anisotropy[38]. Also note that this finite zero-in-plane-field $\Delta H_{sw}/I$ can be suppressed by applying large enough $\pm H_y$, since the activated magnetization would eventually be aligned with the spin polarization direction and results in a zero DL-SOT effective field ($m \times \sigma = 0$).

Finally, we show the switching phase diagrams for both control and tilted anisotropy devices in **Fig.** 5 (c, d). Without tilted anisotropy, the switching phase diagram is symmetric to the origin, and also misses the deterministic switching behavior under field-free condition, as expected. After introducing the laterally tilted anisotropy, a clear shift along the longitudinal field direction is observed, which agrees to that from the hysteresis loop shift measurements and again verifies the existence of a phenomenological longitudinal DL-SOT effective field.

**Conclusion**

In this work, we systematically investigate current-induced field-free SOT switching in Pt/Co/Pt magnetic heterostructures with PMA. Deterministic, robust field-free switching can be achieved in devices with their underlayer Pt prepared by oblique angle deposition, and the switching current can be reduced by tuning the deposition angle. We then verify the mechanism of such field-free switching being magnetic symmetry breaking originated from the laterally tilted anisotropy, which is correlated to the tilted (111) texture of the underlayer Pt. We also demonstrate experimentally and further verify through simulations that the non-linear current dependence of DL torque-induced $\Delta H_{sw}$ is an important feature for such "conventional DL-SOT plus tilted anisotropy" scenario. Additionally, the existence of a longitudinal effective field in the switching phase diagram is another key feature within this scheme, which is resulted from the non-zero longitudinal component of the DL-SOT effective field when activating upon the magnetization. Our work therefore points out the critical role of anisotropy engineering in realizing field-free SOT switching of perpendicular magnetization.

**Acknowledgements**

This study is financially supported by the Ministry of Science and Technology of Taiwan (MOST) under grant No. MOST 111-2636-M-002-012 and by the Center of Atomic Initiative for New Materials (AIMat), National Taiwan University from the Featured Areas Research Center Program within the framework of the Higher Education Sprout Project by the Ministry of Education (MOE) in Taiwan under grant No. NTU-111L9008.

**Conflict of Interest**

All authors have no conflicts of interest.

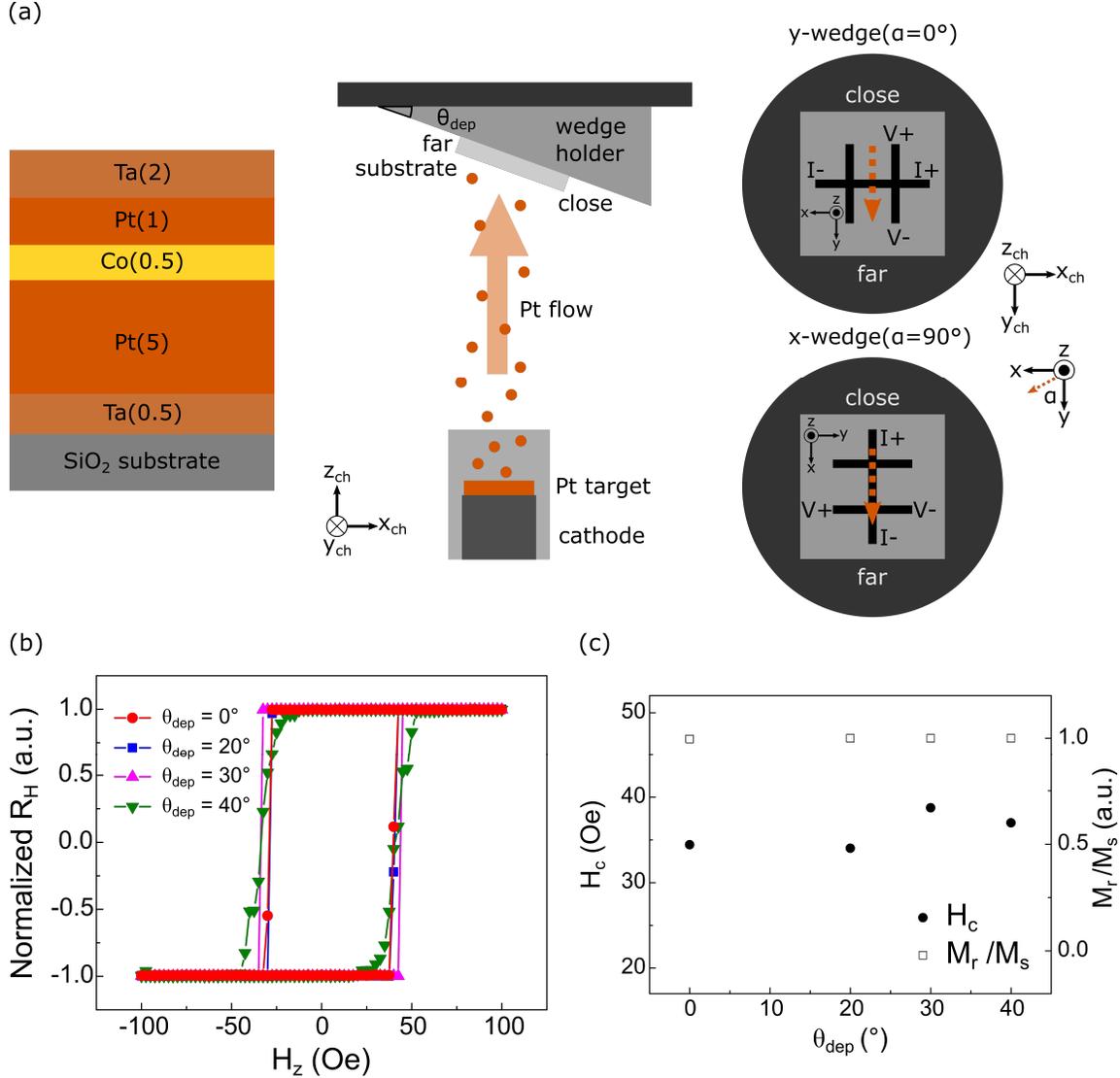

**Fig. 1 Device structures and magnetic properties of Pt(5)/Co(0.5)/Pt(1) prepared by oblique angle deposition. a** Schematic illustrations of the magnetic heterostructures, side-view of the magnetron sputtering chamber and the Hall bar device arrangement on sample holder. The dotted arrow represents the Pt atom flow from the target right below the wedge-shaped holder with tilting angle ($\theta_{dep}$). **b, c** The anomalous Hall effect (AHE) hysteresis loops and the corresponding coercivity ($H_c$) and the degree of perpendicular anisotropy ($M_r/M_s$) of the devices prepared by different $\theta_{dep}$.

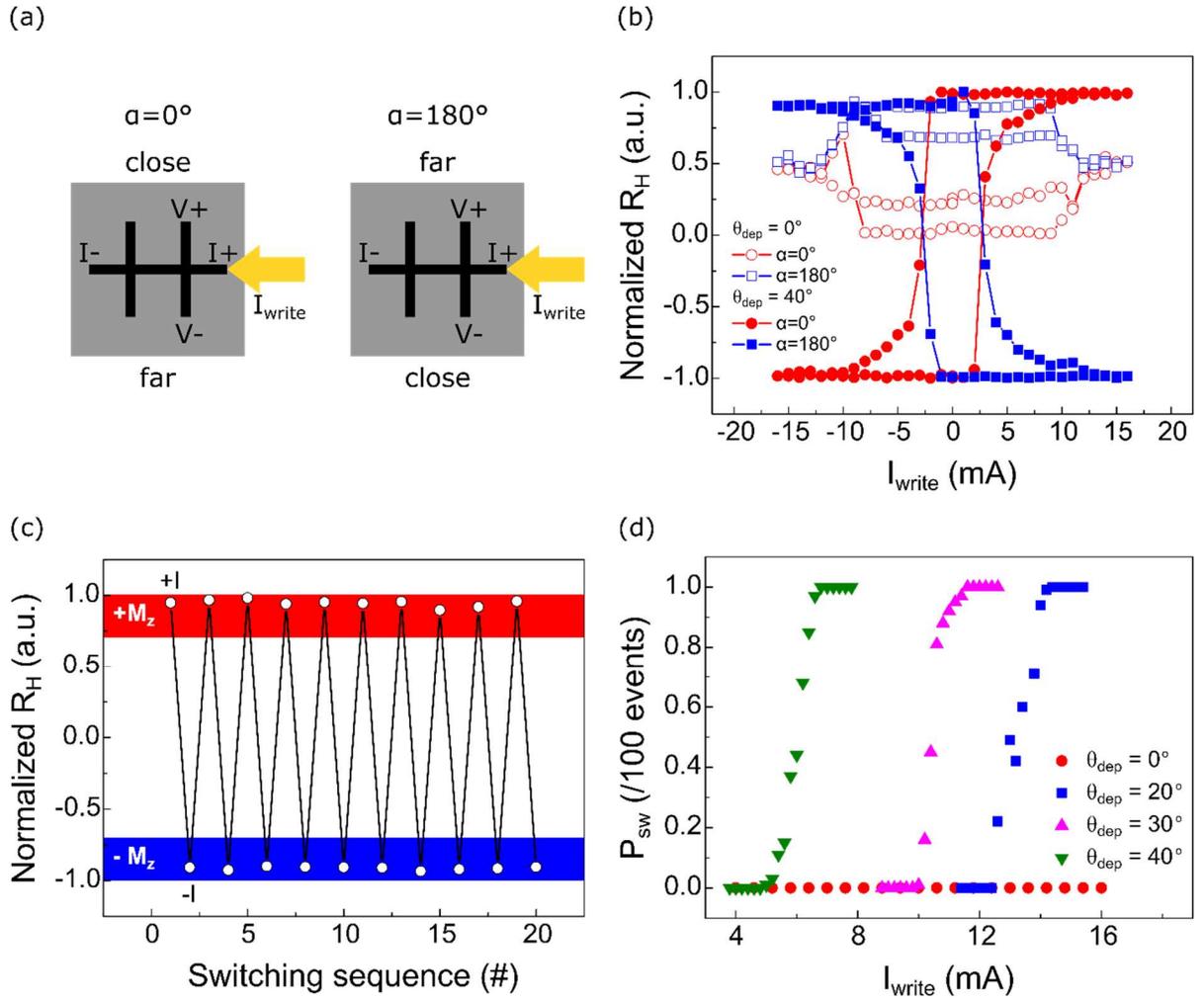

**Fig. 2 Field-free SOT switching of the devices prepared by oblique angle deposition. a** Circuits setup for field-free SOT switching with write current pulse of $I_{\text{write}}$. **b** Representative field-free current-induced switching loops for the device prepared by $\theta_{\text{dep}} = 0°, 40°$ and $\alpha = 0°, 180°$. **c** Representative consecutive switching sequence of a $\theta_{\text{dep}} = 40°$, $\alpha = 0°$ device with $I_{\text{write}}$ set as 7 mA. **d** Switching probability ($P_{sw}$) as a function of $I_{\text{write}}$ for devices prepared under different $\theta_{\text{dep}}$.

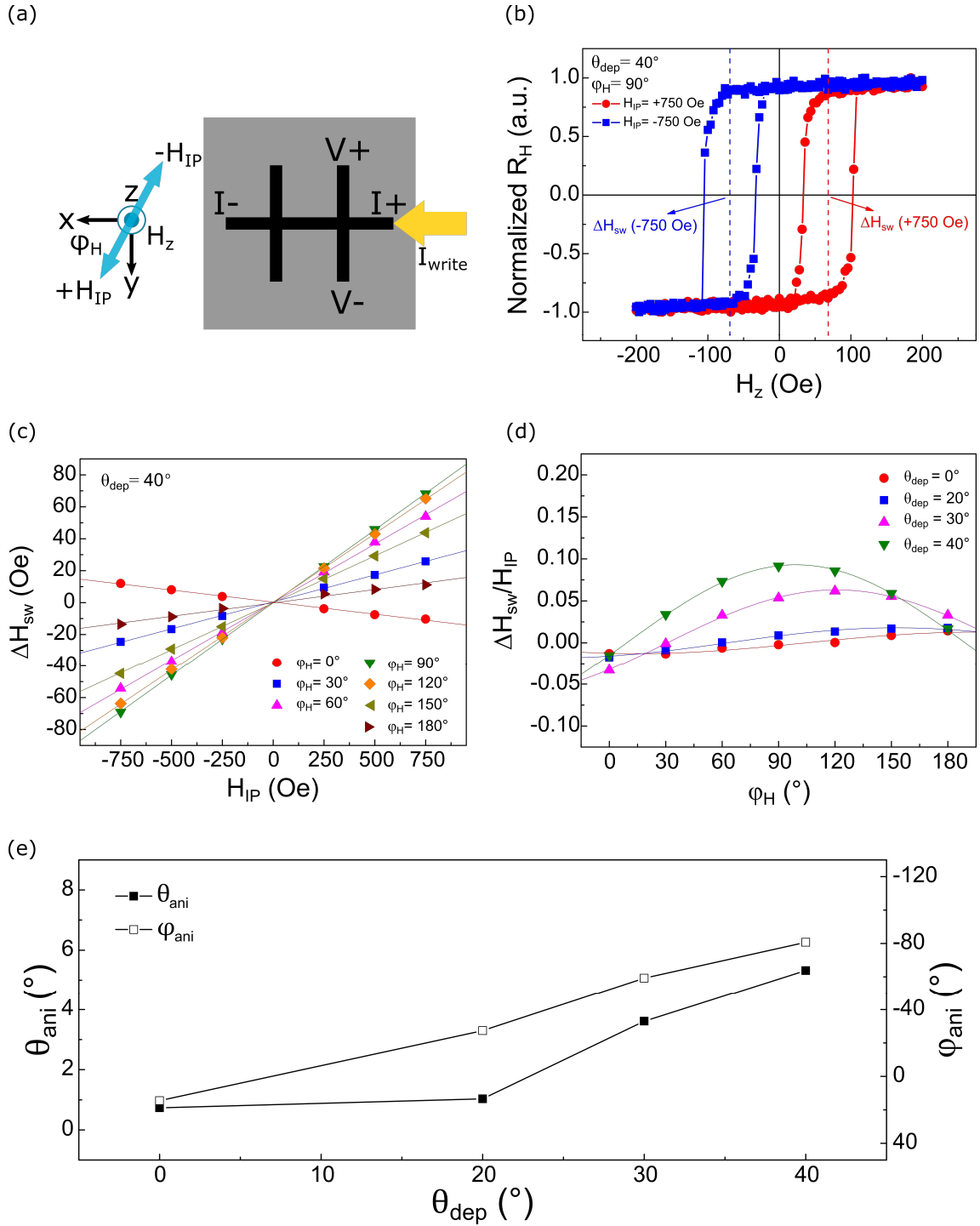

**Fig. 3 Magnetic anisotropy easy axis characterization. a** Experimental setup for AHE hysteresis loop shift measurement with a fixed in-plane field ($H_{IP}$) along a specific azimuthal angle ($\varphi_H$) and scanning out-of-plane field ($H_z$). **b** Representative shift behavior of AHE loops under $H_{IP}$ =

±750 Oe along $\varphi_H = 90°$ and $I_{applied} = 0.1$ mA for a $\theta_{dep} = 40°$ device. **c** The shift of the AHE loop center ($\Delta H_{sw}$) as a function of different $H_{IP}$ with various $\varphi_H$ respect to the $x$-axis. **d** The sinusoidal $\Delta H_{sw}/H_{IP}$ behavior as a function of $\varphi_{IH}$ for devices prepared with different $\theta_{dep}$. **e** Polar angle ($\theta_{ani}$) and azimuthal angle ($\varphi_{ani}$) of the easy axis as functions of $\theta_{dep}$.

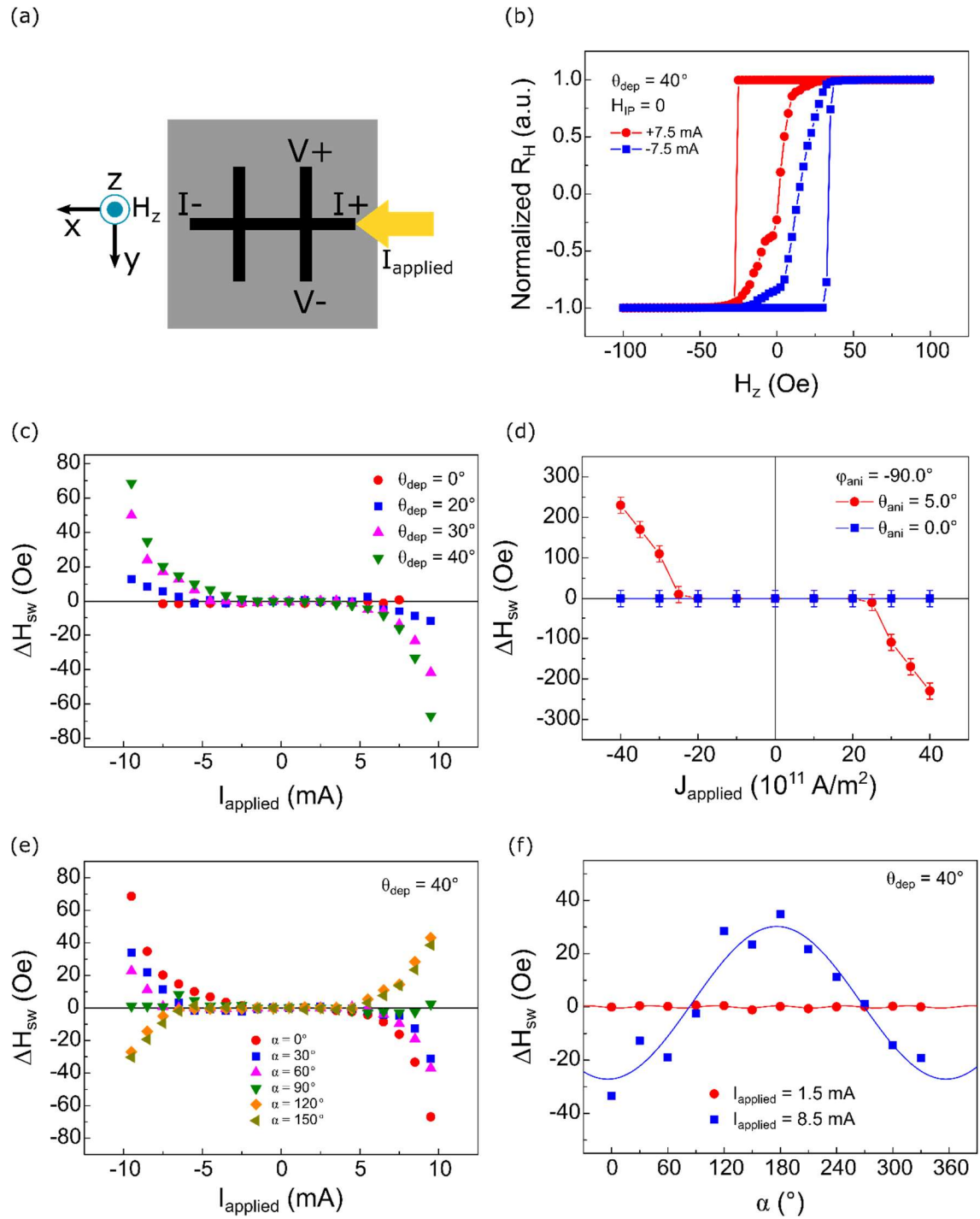

**Fig. 4 Nonlinear damping-like torque response in the presence of tilted anisotropy. a** Experimental setup for AHE hysteresis loop measurements under applying $I_{\text{applied}}$ and zero in-plane external field. **b** Representative AHE loop shifts under $I_{\text{applied}} = \pm 7.5$ mA for a $\theta_{\text{dep}} = 40°$

device. **c** The shifted AHE loop center ($\Delta H_{sw}$) as a function of $I_{applied}$ for devices prepared with different $\theta_{dep}$. **d** Simulated results of $\Delta H_{sw}$ vs. applied current density ($J_{applied}$) for devices with and without laterally tilted anisotropy. The direction of the tilted anisotropy is set as $(\theta_{ani}, \varphi_{ani}) = (5°, -90°)$. **e** $\Delta H_{sw}$ varies with $I_{applied}$ for the $\theta_{dep} = 40°$ devices with different Hall bar arrangements ($\alpha$). **f** $\Delta H_{sw}(1.5\ \text{mA})$ and $\Delta H_{sw}(8.5\ \text{mA})$ as functions of $\alpha$ in a $\theta_{dep} = 40°$ device.

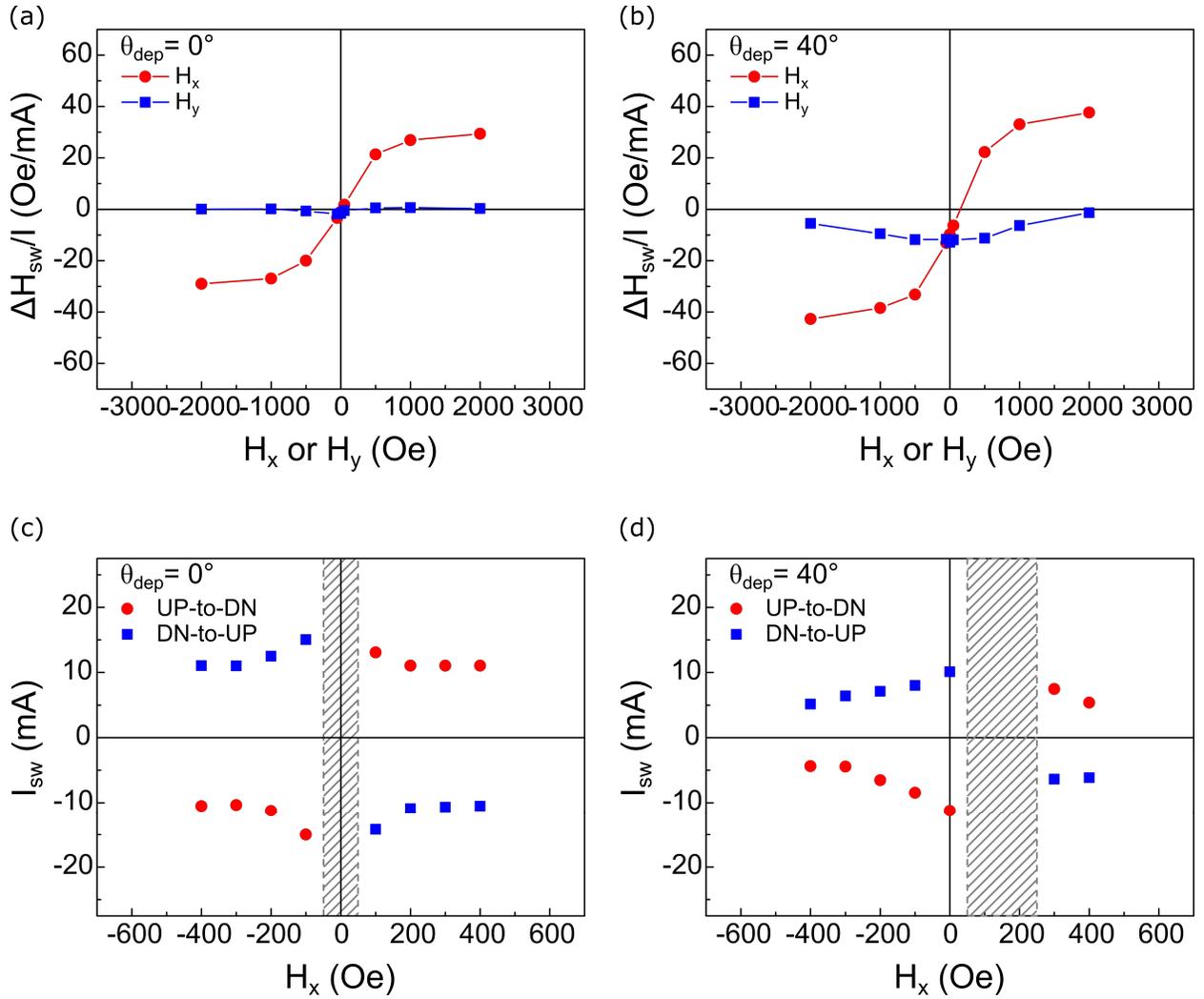

**Fig. 5 Field-dependent SOT properties in devices with and without laterally tilted magnetic anisotropy. a, b** Longitudinal ($H_x$) and transverse ($H_y$) field dependence of DL-SOT-induced loop shift per current ($\Delta H_{sw}/I$) of $\theta_{dep} = 0°$ and $\theta_{dep} = 40°$ devices. The applied current is set at $\pm 6.5$ mA. **c, d** $H_x$ dependence of SOT switching behavior (switching phase diagram) for both types of devices with a channel width of 10 μm. The gray-slash regions correspond to the zones that cannot achieve full switching.